\documentclass[preprint,showpacs,preprintnumbers]{revtex4}
\usepackage[dvips]{graphicx}

\newcommand{\NPB}[3]{Nucl.\ Phys.\ {\bf B#1},\ #2 (#3)}

\newcommand{\PLB}[3]{Phys.\ Lett.\ B\ {\bf #1},\ #2 (#3)}
\newcommand{\PR}[3]{Phys.\ Rep.\ {\bf #1},\ #2 (#3)}
\newcommand{\PRL}[3]{Phys.\ Rev.\ Lett.\ {\bf #1},\ #2 (#3)}

\newcommand{\PRD}[3]{Phys.\ Rev.\ D\ {\bf #1},\ #2 (#3)}
\newcommand{\JPG}[3]{J.\ Phys.\ G\ {\bf #1},\ #2 (#3)}

\newcommand{\ZPC}[3]{Z.\ Phys.\ C\ {\bf #1},\ #2 (#3)}
\newcommand{\EPJC}[3]{Eur.\ Phys.\ J.\ C\ {\bf #1},\ #2 (#3)}

\newcommand{\CPC}[3]{Comp.\ Phys.\ Commun.\ {\bf #1},\ #2 (#3)}
\newcommand{\JHEP}[3]{JHEP\ {\bf #1},\ #2 (#3)}

\newcommand{\ibid}[3]{{\bf #1},\ #2 (#3)}

\begin{document}

\title{Study of color connections in $e^+ e^-$ annihilation}

\author{Feng-lan Shao$^1$}
\email{shaofl@mail.sdu.edu.cn}
\author{Qu-bing Xie$^{2,1}$}
\email{xie@sdu.edu.cn}
\author{Shi-yuan Li$^1$}
\author{Qun Wang$^1$}
\affiliation{$^1$Physics Department, Shandong University, Jinan, 
Shandong 250100, P.R. China }
\affiliation{$^2$China Center of Advanced Science and Technology (CCAST),
Beijing 100080, P. R. China}

\date{\today}

\begin{abstract}

We replace in the event generator JETSET 
the color singlet chain connection 
with the color separate state one as 
the interface between the hard and soft sectors of 
hadronic processes. The modified generator is applied to 
produce the hadronic events in $e^+ e^-$ annihilation. 
It describes the experimental data at the same level as 
the original JETSET with default parameters. This
should be understood as a demonstration that color singlet chain is not the 
unique color connection. 
We also search for the difference in special sets of three-jet events 
arising from different color connections, 
which could subject to further experimental test.

\end{abstract}


\maketitle

\section{Introduction}

Strong interaction processes in high-energy reactions, 
e.g. $e^ + e^ - \to $hadrons, are generally described 
by two distinct sectors: the perturbative and
hadronization phases. The perturbative phase is well
described by perturbative quantum chromodynamics (PQCD),
while the hadronization one is non-perturbative and 
currently can only be described by
hadronization models \cite{Lund}\cite{Webber}. A natural problem 
arises as how to link these two phases, which is beyond 
the capability of PQCD. In present hadronic event generators 
\cite{Jetset}\cite{Herwig}, the color connections are 
assumed to be the color string or the color cluster 
chain. This is true only in the case of $N_C \to \infty $,
because with infinitely many colors 
the probability that two or more partons have 
accidentally the same color is zero, 
and then the way to connect the partons by a string 
or a cluster chain becomes unique \cite{C.Friberg1997}\cite{Q.Wang2000}. 
The present hadronization models work quite well, which shows 
that the large $N_C $ limit does reflect some features of the real 
world. However, with only three colors in nature, 
the color structure of a multiparton state at the end of 
the parton cascade is copious and complex
\cite{Q.Wang2000} - \cite{maltoni}. 
The one that is used in the current string and 
cluster models is called the color singlet chain (CC) state, 
where in each string piece or cluster the color charge from one parton 
is connected to its anti-color from the other, so that 
the string piece or the cluster is color neutral. This 
results in a color neutral flow in the string 
or the cluster system \cite{Q.Wang2001}\cite{maltoni}. 
PQCD calculation shows that, when projected onto 
the color space of the final partonic state, 
the CC state occurs with a probability less than 1 
and the probability decreases as the growing number of partons 
\cite{Q.Wang2001}\cite{Y.Jin}. Part of the rest probability 
goes to such states as the so called color separate (CS) singlet, 
where a multiparton system separates itself into several 
color singlet subsystems and each subsystem 
can hadronize independently
\cite{C.Friberg1997}\cite{Q.Wang2000} 
\cite{T.Sjostrand1994}\cite{L.Lonnblad1996}.

The presence of CS states can be easily 
understood in $e^ + e^ - \to q\bar{q}+ng \to h's $ process. 
If two gluons have opposite colors,
they can form a color singlet subsystem or a closed string. 
When the parton number is large, there are many 
possibilities for two or more gluons to form 
CS states \cite{Q.Wang2000}\cite{Q.Wang1996}.
As is shown in Ref. \cite{Q.Wang2000}, CC and CS states 
are not orthogonal to each other in perturbative sense. 
They belong to two different complete sets of color singlets 
in color space of the final partonic state and are equivalent 
in the context of PQCD. However, they lead to 
different color connections on the hard-soft interface 
and may give rise to different hadronic states through the 
subsequent hadronization process \cite{Q.Wang2000}\cite{Q.Wang2001}. 
Whether the hadronization chooses CC or CS states 
as its starting point is determined by non-perturbative QCD dynamics.
This question become critical in ultra-relativistic heavy ion collisions,
since it is almost impossible to identify a unique color singlet chain in  
a bulk of partons. 

The most efficient and practical way to 
study effects of CS connections on 
hadronic events is to modify the available event generators 
by substituting the CC connection with the CS one. 
Then we can use the new program to 
generate hadronic events and to investigate if 
there are any deviations from those generated with 
the default (CC) connection. 
An explicit example is given in Ref. \cite{Q.Wang2000}, 
where a phenomenological CS model is proposed by 
incorporating CS connection into JETSET. 
The results show that there are 
no significant differences between CS and CC connections 
in terms of global properties of unbiased events \cite{Q.Wang2000}. 
This indicates that the available data for 
these properties does not rule out CS possibility. 
In other words,
color connections beyond the traditional CC one, 
such as the CS connection discussed above, 
are also possible states on the interface between the 
hard and soft sectors in hadronic processes.
To further confirm such an observation, it is natural
to seek new observables in special events which may be sensitive to 
color connections. In this paper, we first give other 
properties of unbiased events except those in \cite{Q.Wang2000} 
and show these two different color 
connections can both describe data. Then we study some properties for 
three-jet events and compare them with 
the recent OPAL \cite{OPAL2002} and 
DELPHI data \cite{DELPHI2000}. 
We find no evidence that CC is the unique color connection on 
the hard-soft interface. However, our investigation 
shows that the CC and CS states 
can lead to larger differences for some selected 
three-jet events where jets are well separated 
and the jet resolution scale, $y_{cut}$, is smaller. This could  
be put to further experimental test, e.g., 
by re-analysis of LEP data or in future Giga-Z experiment 
at linear collider.
 

The outline of this paper is as follows. 
In section 2 we give a brief description of the CS 
model \cite{Q.Wang2000}. We present in section 3
some results about unbiased events. 
In section 4 we study three-jet events 
and propose observables sensitive to color connections. 
A summary of results is given in section 5.

\section{The color separate state model}

Our PQCD analysis shows that up to $O(1/N_C^2)$ a 
partonic system with $q\bar{q}$ and $n$ gluons can be decomposed
into two sub-singlets, one is the CC state made of $q\bar{q}$ 
and $m$ gluons, the other is the CS state made 
of $(n-m)$ gluons \cite{Q.Wang2000}. 
The former forms an open string stretched between the quark and 
the anti-quark spanning $m$ gluons, 
while the latter forms a closed string by the rest gluons. 
These CS states formed in the partonic system 
to $O(1/N_C^2)$ are called leading CS states.
Based on this analysis a phenomenological model is 
constructed in the following way:
(1) Only leading CS states are produced; 
(2) A CS state is produced with the 
relative weight given by the T-measure \cite{B.Andersson1986}. 
We implement the model into a Monte-Carlo program based on JETSET7.4. 
In the program we just replace the default way of color
connection at the end of parton cascade in JETSET7.4, i.e. the CC
connection, with the CS-allowed connection without 
touching all other parts. The CS cluster made of 
two or more gluons in a CS state hadronizes 
as closed strings \cite{Lund}.

Using this program, we have generated CS unbiased events and studied
such global features as thrust, sphericity, oblateness and aplanarity.
They are compared with those from CC unbiased events. 
The comparison shows that there are no significant 
differences between CS and CC unbiased events 
in terms of these global observables. In fact such a
result is not surprising, because for unbiased
events the global properties are mainly determined by PQCD parton
cascade process and are not sensitive to hadronization details. 
This is nothing but the well-known property of local parton-hadron
duality \cite{B.Andersson1989}. 
We therefore need to make further investigation 
in other observables more sensitive to color connections.

\section{Investigation of unbiased events}

\subsection{Identified hadron multiplicities and their momentum spectra}

Two basic observables for hadronic events are
multiplicities of identified hadrons and their momentum distributions. 
Using 6 million Monte Carlo events we 
calculate the average multiplicities and the inclusive 
momentum spectra for $\pi ^\pm $,$K^\pm $,$p(\bar{p})$ 
in unbiased events. The statistical uncertainties 
for these observables are much less than those in data 
(about 1 - 2 order of magnitude lower than data). 
Our results for multiplicities together with data
\cite{Particle2000} at $\sqrt{s}=10GeV, 29GeV$ and $91GeV$ 
are listed in Table 1-3.
The inclusive momentum spectra for $\pi ^\pm$, $K^\pm$, 
$p(\bar{p})$ at same energies are shown in Fig. 1a-c respectively.
The data are taken from Ref.\ \cite{Particle2000}.
The results show that CS unbiased events agree with 
data at the same level as CC ones.

\subsection{Ratio of baryon to meson and baryon-antibaryon correlation}

It is well known that in $e^+e^-$ annihilation 
most of the measured baryons are directly produced.
Even if they are decay products of primary baryons, 
they keep more of their rapidities and
momenta from their parents than most mesons do. 
Therefore, the spectra of the baryons
in particular the correlation between baryon (B) and 
anti-baryon ($\bar{B}$) could be a criteria to test 
different hadronization mechanism \cite{ZG1997}. 
Here we investigate if this correlation
can distinguish different 
color connections before hadronization.

In Lund string fragmentation model, baryon production is described 
by \textit{diquark} or \textit{popcorn} mechanism. 
The diquark production is controlled by the parameter $qq/q$, the ratio 
of the production rate of $qq$ or $\bar{q}\bar{q}$ to that 
of $q\bar{q}$ in color field \cite{Lund}. 
The parameter $qq/q$ controls the ratio of baryon to meson. 
Six million Monte Carlo events 
are produced and the statistical uncertainty is much 
reduced as compared to data. The calculated ratios of baryon to meson at 
$\sqrt{s}=10GeV$, 29GeV and 91GeV are listed in Table 4. 
We see that there is no significant difference between CS and CC connections.
For the CC connection, i.e. default setting in JETSET,
one has to take different values of $qq/q$ at different energies 
to fit data ($qq/q=0.06$ at $\sqrt{s}=10GeV$, 
and $qq/q=0.1$ at other two energies). 
On the other hand, one can use an energy-independent 
value $qq/q=0.1$ for CS events, and the results 
agree with data quite well. 

The popcorn mechanism is realized by the popcorn parameter:
\begin{equation}
\label{eq1}
\rho=\frac{P(BM\bar{B})}{P(B\bar{B})+P(BM\bar{B})},
\end{equation}
where $\rho=1$ means that a meson is always produced 
between B and $\bar{B}$, while in the other limit 
$\rho=0$ it is not at all. 
The popcorn parameter $\rho$ also describes  
the $B\bar{B}$ correlation. Here we take $\rho=0.5$ and 
calculated the $p\bar{p}$, $\Lambda\bar{\Lambda}$ and
$p\bar{\Lambda}/\Lambda\bar{p}$ rapidity
correlations for unbiased events at $Z^0$ pole.  
The results are shown in Fig. 2a-c. 
Both CS and CC connections agree with data \cite{DELPHI1997} at the same level.

Besides the global properties studied in \cite{Q.Wang2000}, 
we have investigated other properties for unbiased events 
with CS or CC color connections. The results obtained in both 
cases are consistent with data. For some results the agreement 
is even better in CS events.  

\section{Investigation of three-jet events at $Z^0$ pole}


Since different color connections lead to different color strings
and may cover different phase space, one expects that 
the CS state should have some effects on the energy and momentum 
flow of final hadrons. So the impact of the CS connection could be observable 
in some specific events. Currently a large number 
of three-jet events, where a gluon-jet can be identified, 
are available at $Z^0$ resonance energy at LEP I. 
It is highly possible that the gluon-jet could be 
the fragmentation product of gluonic color singlet 
clusters which are CS singlets. Therefore, 
we expect that some sensitive observables should be 
out there in a specified window of phase space around the gluon jet. 
In the following, we will study the properties
of three-jet events at $Z^0$ resonance energy
to probe the possible effect of CS states. 
We note that a similar work has been done for 
two-jet events \cite{SY2002}.

\subsection{Selection of three-jet events}

We use Durham jet algorithm \cite{Catani1991}
to select three-jet events. In the algorithm, 
particles are grouped into jets based on a cutoff variable
$y_{cut}$ defined as follows.
For each pair of particles $i$ and $j$ with
energies $E_i$ and $E_j$ respectively and their
opening angle $\theta_{ij}$, one defines $y_{ij}$ as
\begin{equation}
\label{eq2}
y_{ij}\equiv \frac{2\mathrm{min}(E_i^2,E_j^2)
(1-\mathrm{cos}\theta_{ij})}{E_{vis}^2},
\end{equation}
where $E_{vis}$ is the total visible energy in the event.
Given a cutoff value $y_{cut}$,
one can put all particles into one jet
where all $y_{ij}$'s for any two particles inside the jet are
smaller than $y_{cut}$, while $y_{ij}>y_{cut}$
for the case with one inside the jet and the other outside it.
The number of particles in a jet
is then the multiplicity of the jet. Increasing
$y_{cut}$ makes the fraction of multi-jet events lower
because $y_{cut}$ is actually a jet resolution parameter.

We have calculated the fraction of three-jet events 
as function of $y_{cut}$. Both color connections agree 
with data\cite{P.D.Acton1992}.
We have also calculated the charged particle multiplicity
for three-jet events at some values of $y_{cut}$ and 
compared our results with data \cite{DELPHI1992}.
There is no sizable difference between CS and CC events.

\subsection{Particle multiplicities and momentum spectra of 
identified hadrons in Y events}

Recently OPAL collaboration has measured the charged particle
multiplicity $N_{ch}$ for three-jet light-quark 
events of the so-called Y shape from $Z^0$ decay 
as function of the angle $\theta$ between 
the two jets with lowest energies \cite{OPAL2002}.
The jets are defined by Durham jet-finder.
The resolution $y_{cut}$ is adjusted separately for
each tagged $uds$ event so that three jets can be exactly 
reconstructed. They applied this procedure instead of using 
a fixed resolution value just to avoid introducing a bias 
into the gluon jet. The jets are labelled 1,2 and 3 
in such an order that jet 1 carries the highest energy.
Only those events, the so-called \textit{Y events}, 
are kept where the angles between the most energetic 
jet and the rest two are the same up to $3^{\circ}$. 
We are particularly interested in these data which could be used to 
test our CS model. We choose such Y events from 10 million 
Monte Carlo samples at $Z^0$ pole accommodating the same 
experimental condition. The calculated particle multiplicities
varied with angle $\theta$ are shown in Fig. 3. 
One can see that there is some difference between 
two types of color connections and that results from the 
CC connection seem to agree with data better 
than those from the CS one. However, 
such a difference is not significant enough to 
discriminate the two in terms of data.
The multiplicity distribution calculated for these Y events 
with $0^{\circ}< \theta <{120^{\circ}}$ is shown in Fig. 4.  
Besides these results, we also calculated other observables measured 
in Ref. \cite{OPAL2002}. All these results fail to show
any significant difference between the two types of connections.

DELPHI collaboration has measured momentum spectra of identified hadrons
in quark jets of all species ($duscb$) together with gluon jets 
for Y events\cite{DELPHI2000}. The three-jet events are 
identified by Durham algorithm with $y_{cut}=0.015$. 
The jets are also labelled in descending order of jet energies. 
The Y events are obtained with $\theta_{2},\theta_{3} 
\in [150^\circ-15^\circ,150^\circ+15^\circ]$, where $\theta_{2}$ and 
$\theta_{3}$ are angles between the highest energetic  
jet and other two jets with lower energies. 
In order to compare with data of this experiment, 
we calculate the momentum spectra of identified hadrons in quark and gluon
jets for Y events with the same constraints as in the experiment.  
The samples are 10 million Monte Carlo events at $ Z^0 $ pole. 
The results are shown in Fig. 5. One sees that it is hard to 
distinguish the CS from the CC connection by comparison with data.
So up to now,  no evidence has ruled out the CS connection as 
a possible candidate state on the hard-soft interface.

However, the above results imply that some differences between 
the CC and CS connections are present. In the next section we will 
investigate whether they could be amplified 
by choosing suitable $y_{cut}$ and the angles between jets.

\subsection{Observables in special three-jet events }

One expects that more sensitive observables come from
the phase space around the gluon jet. 
A normal method to identify the gluon-jet 
in experiments is to label three jets by their energies 
in descending order $E_1\geq E_2\geq E_3$.
The most energetic jet (jet 1) is usually recognised
as the quark-jet (or antiquark-jet),
and the one with the smallest energy (jet 3) as the gluon jet.
We denote the angle between the first and the second jet
by $\theta_{12}$ and that between the first and the
third one by $\theta_{13}$.

We look at three-jet events
with restricted range of $\theta_{12}$ and $\theta_{13}$.
We calculate charged particle multiplicities 
whose sensitivity to different color connections
varies with $\theta_{12}$ and $\theta_{13}$.
We find that only those events where 
jet 3 is well separated from jet 1 and 2 
exhibit considerable difference in multiplicity from color connections.
Considering both the statistics 
and the sensitivity to color connection, we choose
${100^{\circ}}< \theta_{12}, \theta_{13}< {160^{\circ}}$.
We also find that the difference increases with decreasing $y_{cut}$.
Here we use a fixed value of the resolution parameter $y_{cut}$ 
instead of one for each three-jet event.

Now we focus on the gluon jet in three-jet events.
We calculate the multiplicity, 
the invariant mass and the longitudinal
momentum of the gluon jet as functions of $y_{cut}$. 
Compared with those from CC connections, 
the results from CS ones are found to be smaller for the multiplicity
and the invariant mass, and larger for the longitudinal momentum.
We also find that the difference between CC and CS
is more significant for smaller $y_{cut}$, as shown in Fig.6.
We select three-jet events from 10 million Monte Carlo samples 
at $Z^0$ pole. Our numerical results are given with  constraint 
${100^{\circ}}< \theta_{12}, \theta_{13}< {160^{\circ}}$ 
at $y_{cut}=0.0005$. The fraction of selected three-jet events 
in the total ones is about $0.502\%$. We give in Fig. 7 
the multiplicity distribution. 
The result shows that two different ways of color connections 
do bring larger differences in the final hadronic state. 
The shapes of two multiplicity distributions
look similar, but the peak of the distribution
for the CS case locates at a smaller value of $n_{ch}$ than
that for the CC case, implying a lower average multiplicity:
${\langle n_{ch} \rangle}_{CS} =18.22$, 
${\langle n_{ch}\rangle}_{CC} =19.41$, and 
the difference is $\triangle \langle n_{ch} \rangle =1.19 $.
The reason is obvious: in CS events several groups of gluons form
color singlet clusters and each hadronizes independently, which
makes the effective energy to produce hadrons smaller than
that in CC events.

Considering that the average longitudinal
momentum of the gluon-jet for CS events
is larger than that for CC ones, 
we define the following observable:
\begin{equation}
\label{eq3}
T=\frac{\sum\limits_{i}p_{zi}}{\sum\limits_{i}|\mathbf{p}_i|},
\end{equation}
where $\mathbf{p}_i$ and $p_{zi}$ are the 3-momentum and its
longitudinal component of a final state hadron in the gluon-jet respectively.
Here \textit{longitudinal} means parallel to the gluon-jet axis.
The observable $T$ is introduced to amplify the difference of CC and CS
events. The distribution of observable $T$ with both CS and CC
connections is given in Fig. (8a) and shows that T is a sensitive 
observable to color connections. We also calculated the spectra of the
charged particle multiplicity, the rapidity and the invariant mass in the
gluon-jet, as shown in Fig.(8b-d). One sees that now
the predicted distinction between CC and CS connections is obvious. 
The reason is that the formation of closed
strings in a gluon jet with the CS connection 
takes more momentum along the
the direction of the mother parton, i.e. the gluon,  
than the CC one, which makes the gluon-jet thinner. 
This point can be further strengthened by the spectra of the polar angle
and the longitudinal momentum for hadrons in the gluon jet.
We note that all results in Fig.8 are consistent 
with each other.

\section{Summary}

In this paper, the comparison of our results with the available data 
up to now shows that 
the CS state is a possible choice for the hard-soft interface 
in hadronic processes. This is consistent with PQCD 
analysis. Such a picture of colour connection is natural 
for a deconfined quark-gluon system, where a huge number of 
partons interact with each other and it is impossible to 
identify a unique color singlet chain. 
This paper is also an attempt for searching
observables in special events where
the differences between the CS and CC states are significant. 
The identification of these differences in experiments 
will deepen our understanding of the hard-soft interface 
in hadronic processes.

By a careful look at certain type of three-jet events with smaller
$y_{cut}$,
where angles between jets are restricted                                          
in a specific range, we have found
more sensitive observables to color connections.
The difference between CC and CS connections is amplified 
in the charged particle multiplicity and properties of the gluon-jet
in these events. The distinction between CS and CC events 
can be put to test by re-analysing the  LET-I data 
and/or in future experiments such as Giga-Z at linear collider.

\subsection*{Acknowledgements}

The authors thank Z.-T. Liang and Z.-G. Si for insightful discussions.
This work is supported in part by the
National Natural Science Foundation of China under
grant No. 10075031 and 10205009.

\newpage


\begin{table}
\caption{Average multiplicities for hardons in  
$e^+e^-$ annihilation at $\sqrt{s}=10GeV $. 
For each entry, the charge conjugate state is also 
included if it is different from the entry itself.}

\begin{tabular}{cccc}
\hline
particle & data & CC & CS \\
\hline
$\pi^+$  & $6.6 \pm 0.2$ & 6.12 & 6.31\\
$\pi^0$ & $3.2 \pm 0.3$ & 3.59 & 3.71\\
$\eta$ & $0.2 \pm 0.04$ & 0.34 & 0.37\\
$\rho(770)^0$ & $0.35 \pm 0.04$ & 0.50 & 0.53\\
$\omega$ & $0.30 \pm 0.08$ & 0.41 & 0.44\\
$\eta'$ & $0.03 \pm 0.01$ & 0.095 & 0.099\\
$\phi(1020)$ & $0.044 \pm 0.003$ & 0.073 & 0.085\\
$K^+$ & $0.90 \pm 0.04$ & 0.98 & 0.99\\
$K^0$ & $0.91 \pm 0.05$ & 0.85 & 0.85\\
$K^*(892)^+$ & $0.27 \pm 0.03$ & 0.42 & 0.41\\
$K^*(892)^0$ & $0.29\pm 0.03$ & 0.37 & 0.36\\
$D^+$ & $0.16 \pm 0.03$ & 0.17 & 0.17\\
$D^0$ & $0.37 \pm 0.06$ & 0.48 & 0.48\\
$D^*(2012)^+$ & $0.22 \pm 0.04$ & 0.24 & 0.24\\
$D_s^+$ & $0.13 \pm 0.02$ & 0.097 & 0.097\\
$p$ & $0.253 \pm 0.016$ & 0.342 & 0.298\\
$\Delta(1232)^{++}$ & $0.040 \pm 0.010$ & 0.051 & 0.041\\
$\Lambda$ & $0.080 \pm 0.007$ & 0.107 & 0.093\\
$\Sigma^0$ & $0.023 \pm 0.008$ & 0.021 & 0.019\\
$\Xi^-$ & $0.0059 \pm 0.0007$ & 0.0074 & 0.0065\\
$\Sigma(1385)^ \pm$ & $0.010  \pm 0.0020$ & 0.0172 & 0.0136\\
$\Xi(1530)^0$ & $0.0015 \pm 0.0006$ & 0.0013 & 0.0011\\
$\Lambda_c^+$ & $0.100 \pm 0.030$ & 0.049 & 0.044\\
\hline
\end{tabular}
\end{table}

\begin{table}
\caption{Same as Table 1 except at  $\sqrt{s}=29GeV$}


\begin{tabular}{cccc}
\hline
particle & data & CC & CS \\
\hline
$\pi^+$  & $10.3 \pm 0.4$ & 10.6 & 11.0\\
$\pi^0$ & $5.83 \pm 0.28$ & 6.11 & 6.32\\
$\eta$ & $0.61 \pm 0.07$ & 0.61 & 0.66\\
$\rho(770)^0$ & $0.81 \pm 0.08 $ & 0.91 & 0.96\\
$\omega$ & $ $ & 0.79 & 0.83\\
$\eta'$ & $0.26 \pm 0.10$ & 0.18 & 0.18\\
$\phi(1020)$ & $0.085 \pm 0.011$ & 0.117 & 0.132\\
$K^+$ & $1.48 \pm 0.09$ & 1.52 & 1.54\\
$K^0$ & $1.48 \pm 0.07$ & 1.34 & 1.35\\
$K^*(892)^+$ & $0.64 \pm 0.05$ & 0.69 & 0.68\\
$D^+$ & $0.17 \pm 0.03$ & 0.19 & 0.19\\
$D^0$ & $0.45 \pm 0.07$ & 0.55 & 0.55\\
$D^*(2012)^+$ & $0.43 \pm 0.07$ & 0.27 & 0.27\\
$D_s^+$ & $0.45 \pm 0.20$ & 0.12 & 0.12\\
$p$ & $0.640 \pm 0.050$ & 0.694 & 0.631\\
$\Delta(1232)^{++}$ & $$ & 0.109 & 0.094\\
$\Lambda$ & $0.205 \pm 0.010$ & 0.214 & 0.193\\
$\Sigma^0$ & $0.023 \pm 0.008$ & 0.021 & 0.019\\
$\Xi^-$ & $0.0176 \pm 0.0027$ & 0.0150 & 0.0133\\
$\Sigma(1385)^ \pm$ & $0.033  \pm 0.008$ & 0.038 & 0.032\\
$\Xi(1530)^0$ & $$ & 0.0027 & 0.0022\\
$\Lambda_c^+$ & $0.110\pm 0.050$ & 0.068 & 0.067\\
\hline
\end{tabular}
\end{table}

\begin{table}
\caption{Same as Table 1 except at $\sqrt{s}=91GeV $}


\begin{tabular}{cccc}
\hline
particle & data & CC & CS \\
\hline
$\pi^+$  & $16.99 \pm 0.27$ & 16.94 & 17.06\\
$\pi^0$ & $9.47 \pm 0.54$ & 9.58 & 9.69\\
$\eta$ & $0.971 \pm 0.030$ & 1.003 & 1.043\\
$\rho(770)^0$ & $1.231 \pm 0.098 $ & 1.503 & 1.522\\
$\omega$ & $1.08 \pm 0.12 $ & 1.35 & 1.36\\
$\eta'$ & $0.156 \pm 0.021$ & 0.297 & 0.295\\
$\phi(1020)$ & $0.0963 \pm 0.0032$ & 0.1932 & 0.2077\\
$K^+$ & $2.242 \pm 0.063$ & 2.300 & 2.305\\
$K^0$ & $2.013 \pm 0.033$ & 2.070 & 2.058\\
$K^*(892)^+$ & $0.715 \pm 0.059$ & 1.102 & 1.074\\
$K^*(892)^0$ & $0.738\pm 0.024$ & 1.096 & 1.069\\
$D^+$ & $0.175 \pm 0.016$ & 0.175 & 0.174\\
$D^0$ & $0.454 \pm 0.030$ & 0.489 & 0.489\\
$D^*(2012)^+$ & $0.183 \pm 0.010$ & 0.240 & 0.270\\
$D_s^+$ & $0.131 \pm 0.21$ & 0.130 & 0.130\\
$p$ & $1.048 \pm 0.045$ & 1.195 & 1.090\\
$\Delta(1232)^{++}$ & $0.085 \pm 0.014$ & 0.188 & 0.164\\
$\Lambda$ & $0.374 \pm 0.009$ & 0.385 & 0.351\\
$\Sigma^0$ & $0.070 \pm 0.012$ & 0.073 & 0.067\\
$\Sigma^ \pm $ & $0.174 \pm 0.009$ & 0.140 & 0.127\\
$\Xi^-$ & $0.0258 \pm 0.0010$ & 0.0274 & 0.0248\\
$\Sigma(1385)^ \pm$ & $0.0462  \pm 0.0028$ & 0.0738 & 0.0652\\
$\Xi(1530)^0$ & $0.0055 \pm 0.0005$ & 0.0054 & 0.0048\\
$\Lambda_c^+$ & $0.078\pm 0.017$ & 0.059 & 0.059\\
\hline
\end{tabular}
\end{table}

\begin{table}
\caption{ The ratios of baryon to meson. We set $qq/q=0.1$.
The data are from Ref.\ \cite{Particle2000}.
The numbers marked with `` * " are those beyond two standard
deviations from data. }


\begin{tabular}{ccccc}
\hline
$\sqrt{s}/GeV$ &particle & data & CC & CS \\
\hline
$ $ & $p/\pi^+$ & $0.062 \pm 0.003$ & $0.071^*$ & 0.064\\
$ $ & $p/K^+$   & $0.467 \pm 0.024$  & $0.520^*$ & 0.473\\
$91$ & $\Lambda/K^+$ & $0.167 \pm 0.006$ & 0.167 & $0.152^*$\\
$ $ & $\Sigma(1385)^\pm/K^{*+}$ & $0.0646 \pm 0.0066$  & 0.0670 & 0.0607\\
$ $ & $\Xi(1530)^0/K^{*+}$   & $0.0075 \pm 0.0007$  & $0.0049^*$ &
$0.0045^*$\\
\hline
$ $ & $p/\pi^+$ & $0.062 \pm 0.005$ & 0.065 & 0.057\\
$ $ & $p/K^+$   & $0.432 \pm 0.043$  & 0.453 & 0.407\\
$29$ & $\Lambda/K^+$ & $0.139 \pm 0.011$ & 0.141 & 0.126\\
$ $ & $\Sigma(1385)^\pm/K^{*+}$ & $0.0516 \pm 0.0131$  & 0.0553 & 0.0468\\
$ $ & $\Xi(1530)^0/K^{*+}$   & $ $  & 0.0045 & 0.0038\\
\hline
$ $ & $p/\pi^+$ & $0.038 \pm 0.003$ & $0.056^*$ & $0.047^*$\\
$ $ & $p/K^+$   & $0.281 \pm 0.022$  & $0.349^*$ & 0.301\\
$10$ & $\Lambda/K^+$ & $0.089 \pm 0.009$ & $0.109^*$ & 0.094\\
$ $ & $\Sigma(1385)^\pm/K^{*+}$ & $0.0393 \pm 0.0086$  & 0.0410 & 0.0332\\
$ $ & $\Xi(1530)^0/K^{*+}$   & $0.0052 \pm 0.0021$  & 0.0035 & 0.0031\\
\hline

\end{tabular}

\end{table}




\begin{figure}

\includegraphics[scale=0.4]{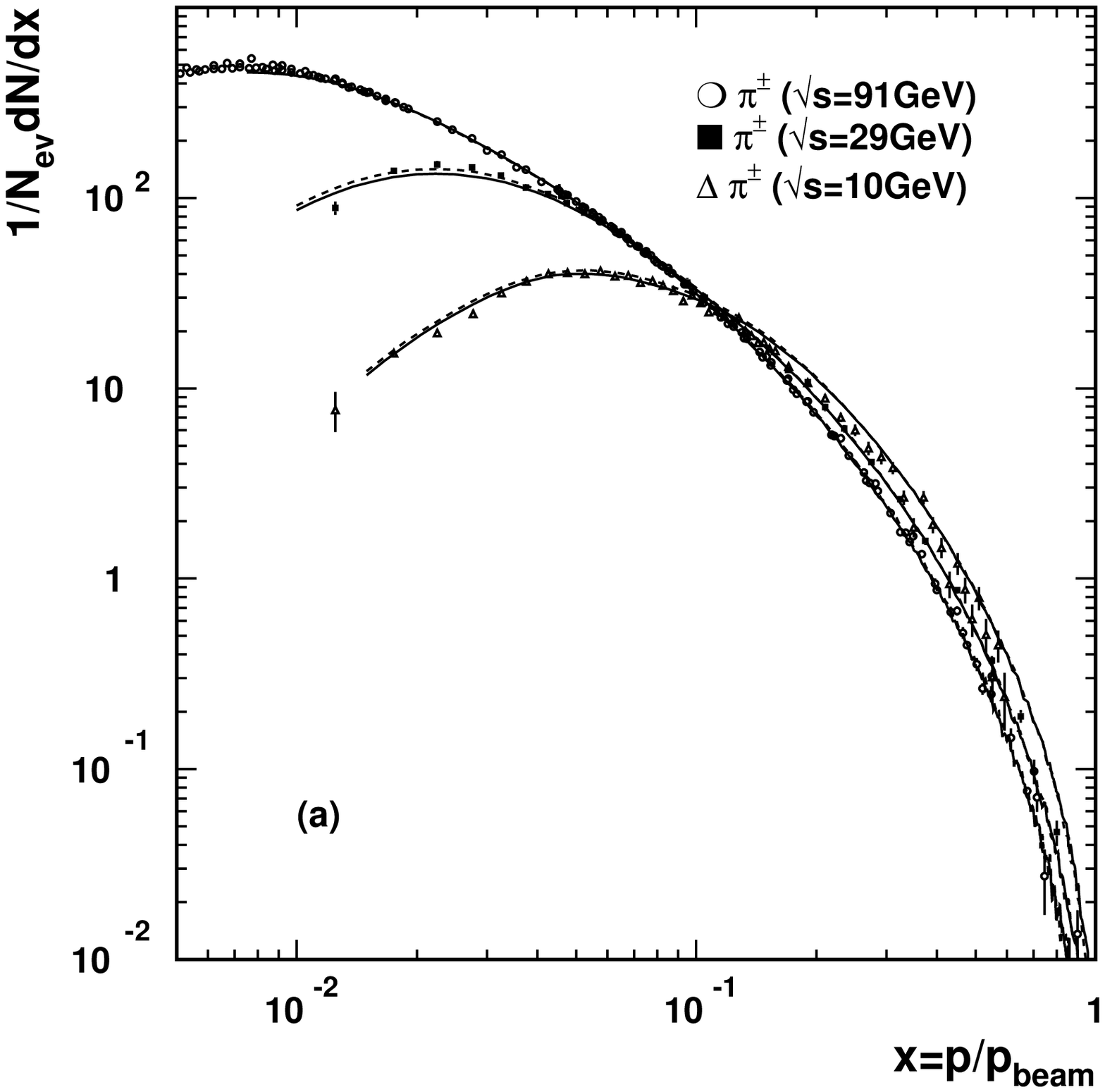}
\includegraphics[scale=0.4]{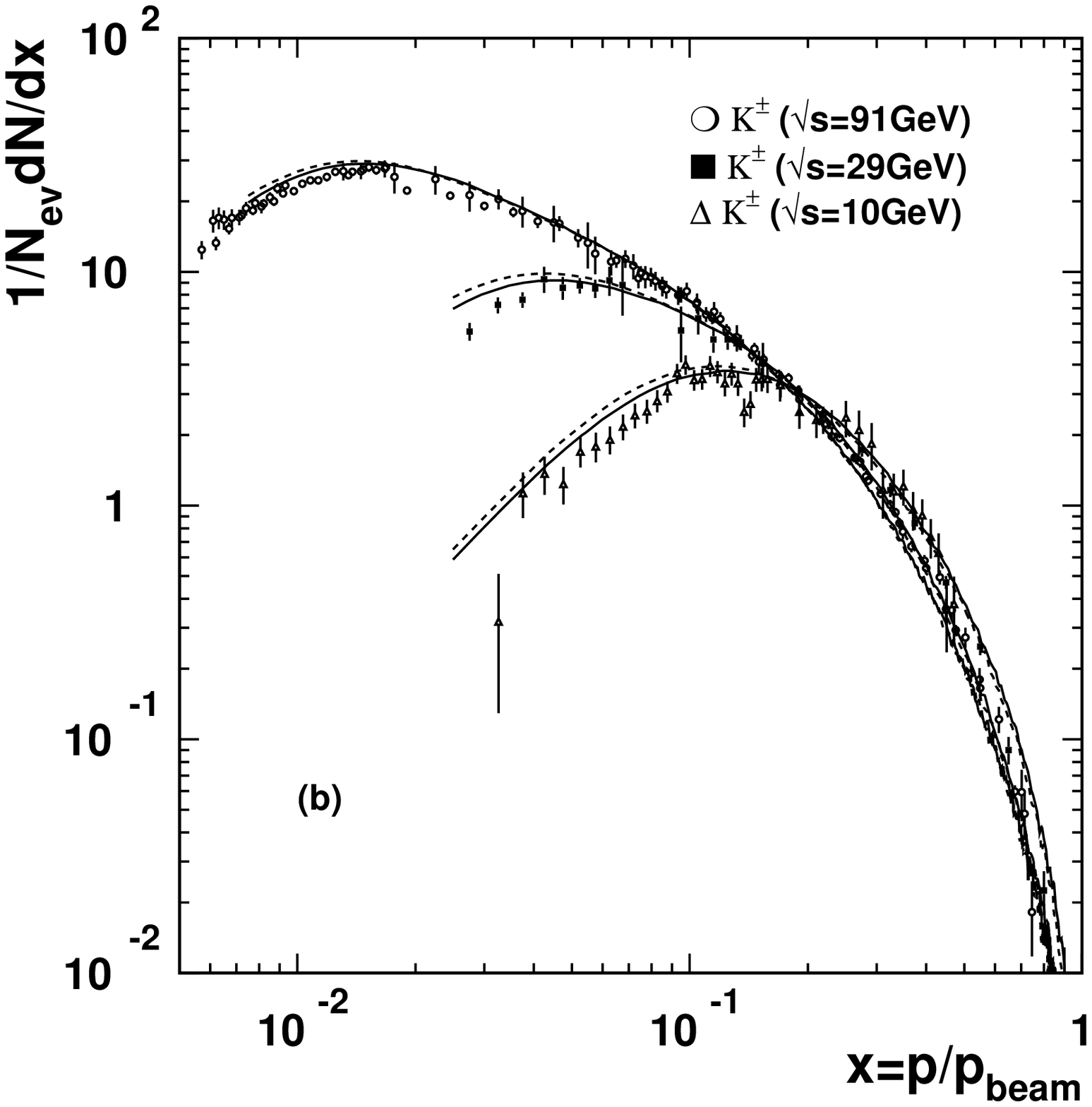}
\includegraphics[scale=0.4]{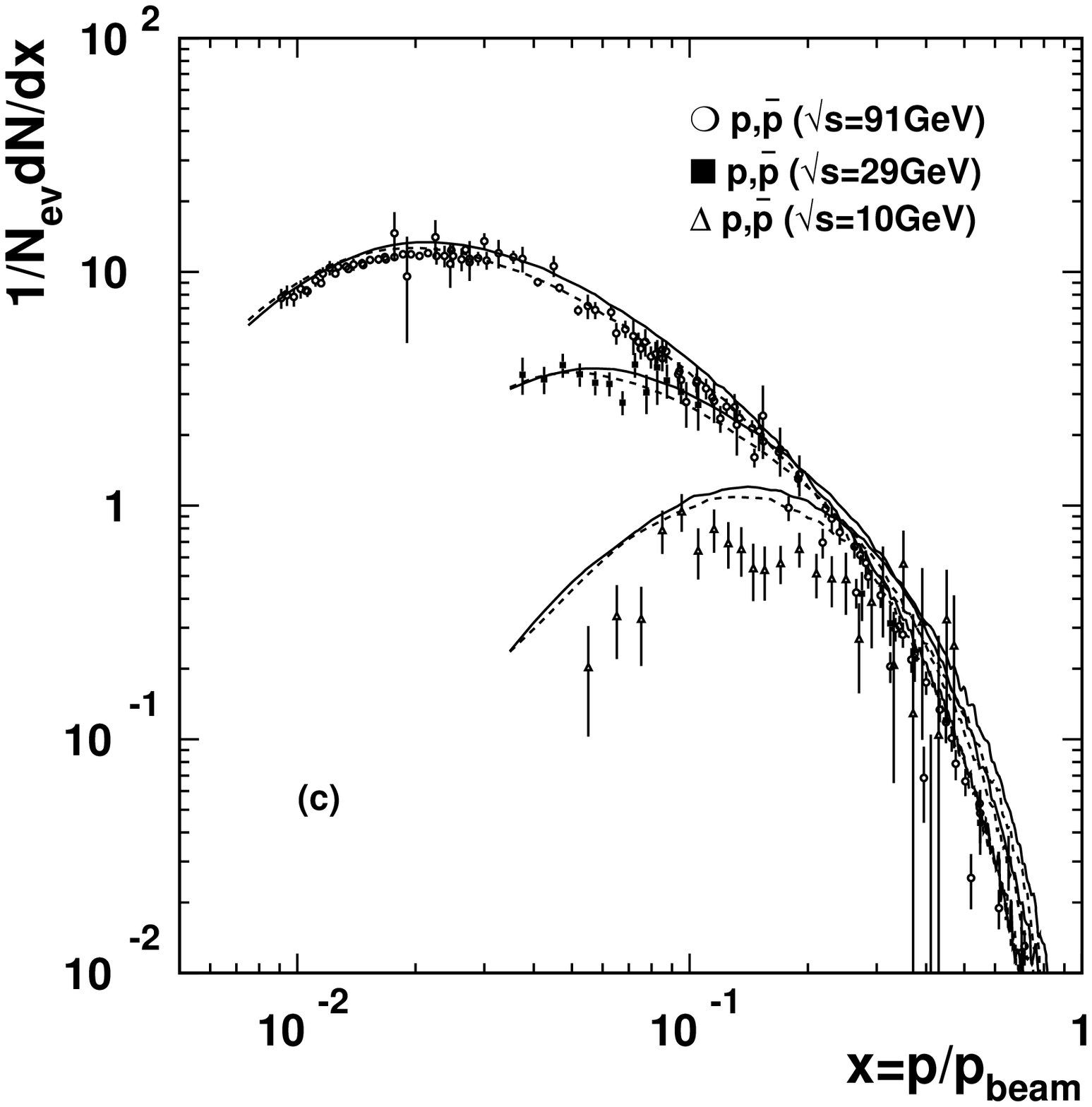}

\caption{
Inclusive cross sections $(1/N_{ev})(dN/dx)$ for (a)  $\pi^ \pm$,
(b) $K^\pm$ and (c) $p\bar{p}$ as function of $x=p/p_{beam}$.
The solid lines are from CC events and the dashed ones from CS
events. The data are taken from Ref.\ \cite{Particle2000}.
}
\label{figs(1a)-(1c)}

\end{figure}

\begin{figure}


\includegraphics[scale=0.4]{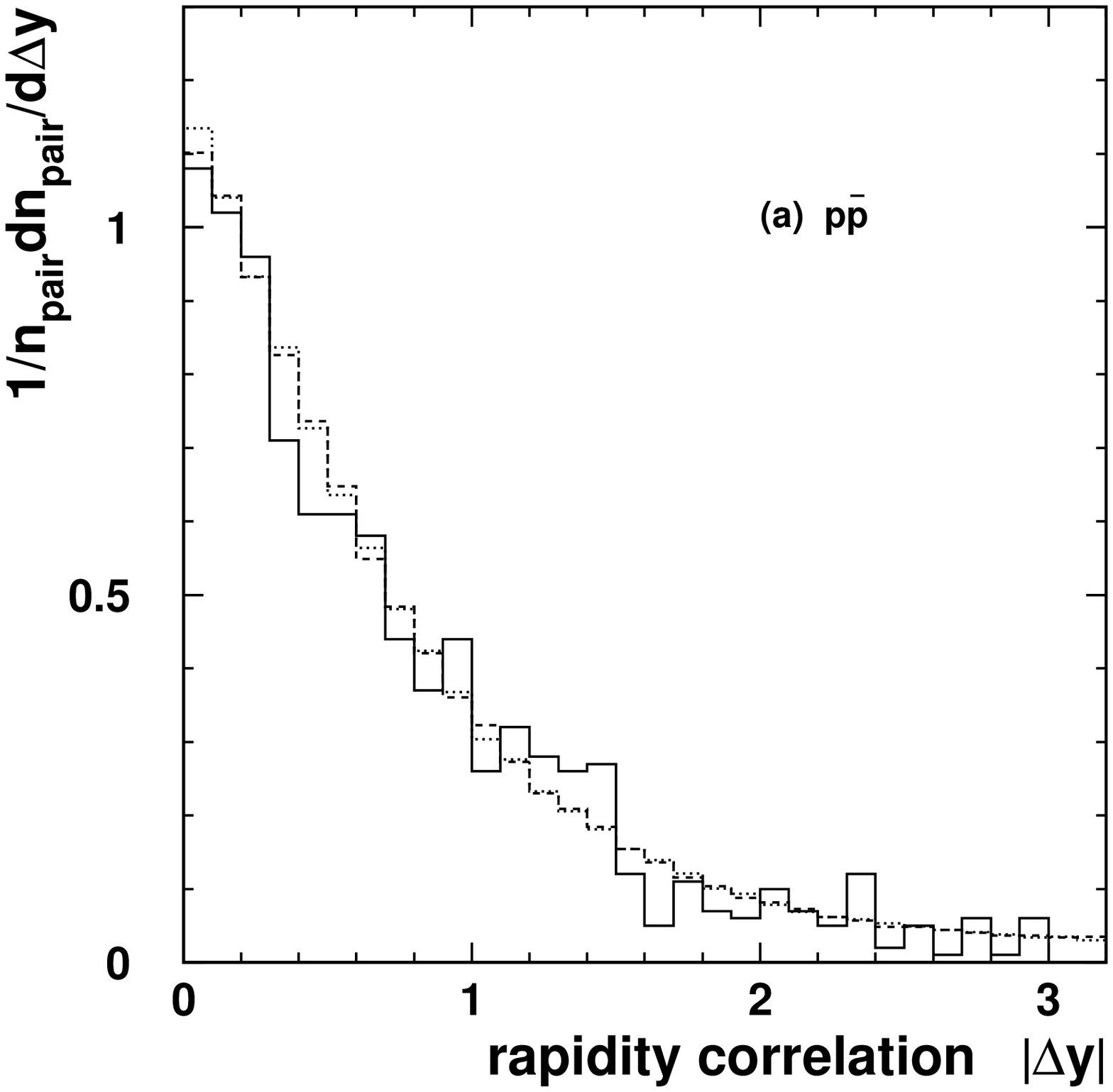}
\includegraphics[scale=0.4]{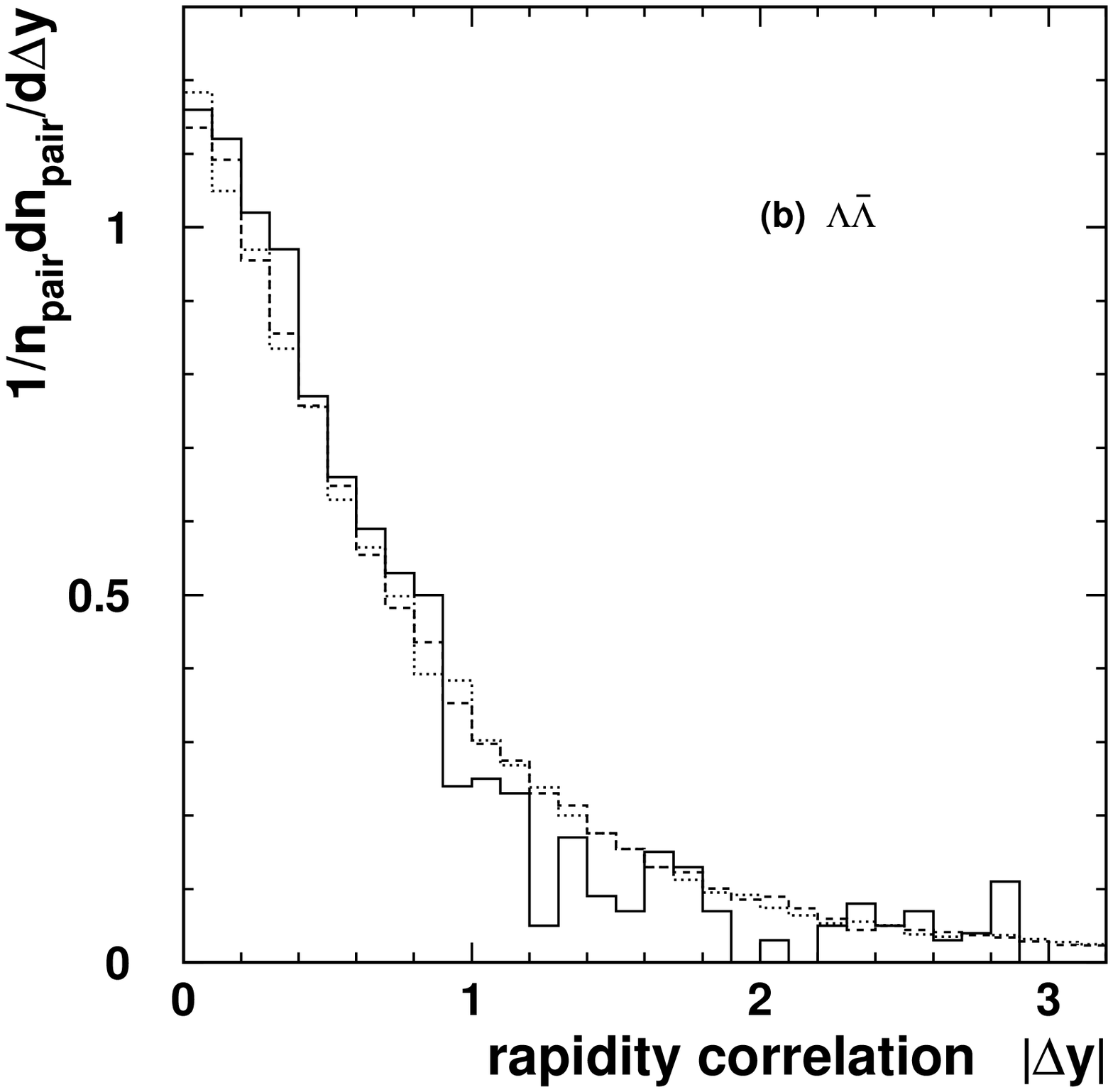}
\includegraphics[scale=0.4]{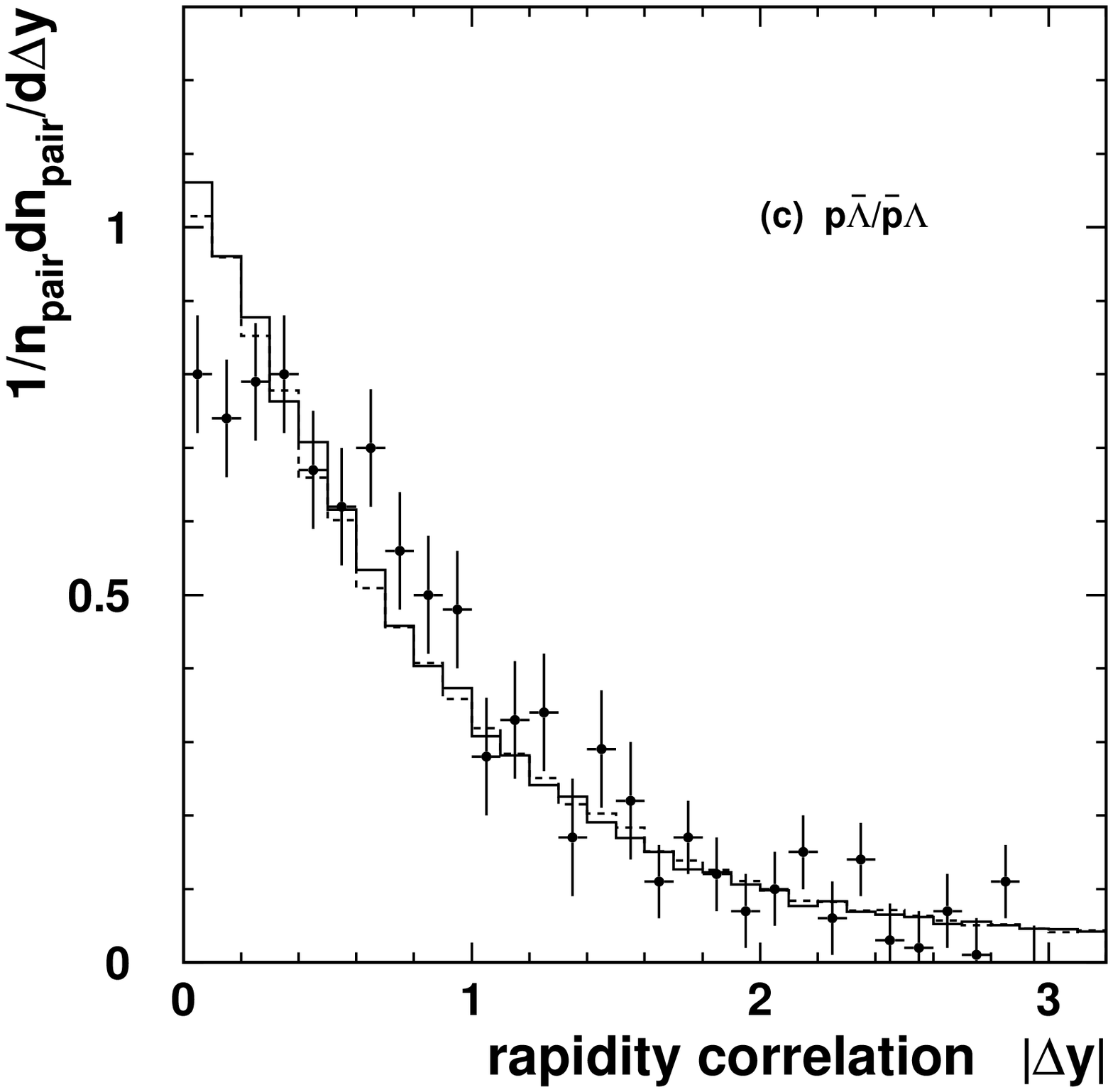}

\caption{Rapidity differences of $B\bar{B}$ pairs with respect to the thrust
axis of the event.
In (a) and (b), the solid histograms are data from Ref. \cite{DELPHI1997};
The dashed and dotted lines are from CC and CS events respectively;
In (c), the solid and dashed lines are from CC and CS events respectively;
The black squares with error bars are data \cite{DELPHI1997}.
}
\label{figs.(2a)-(2c)}
\end{figure}

\begin{figure}

\includegraphics[scale=0.8]{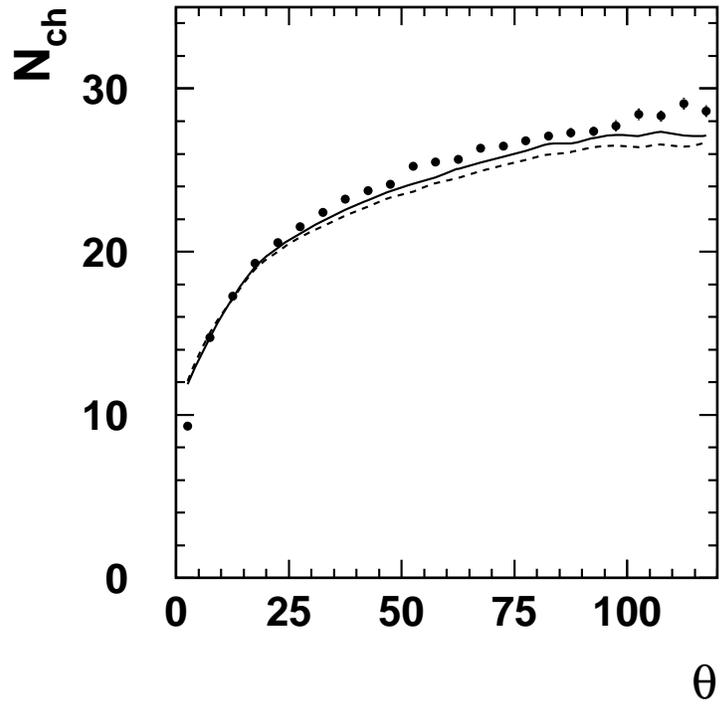}

\caption{The particle multiplicity $n_{ch}$ 
for Y events as function of the angle $\theta$ 
between two jets of the lowest energies.
The solid line is from CC events and the dashed one
from CS events; The data are taken from Ref. \cite{OPAL2002}.
}

\label{fig.3}
\end{figure}

\begin{figure}

\includegraphics[scale=0.8]{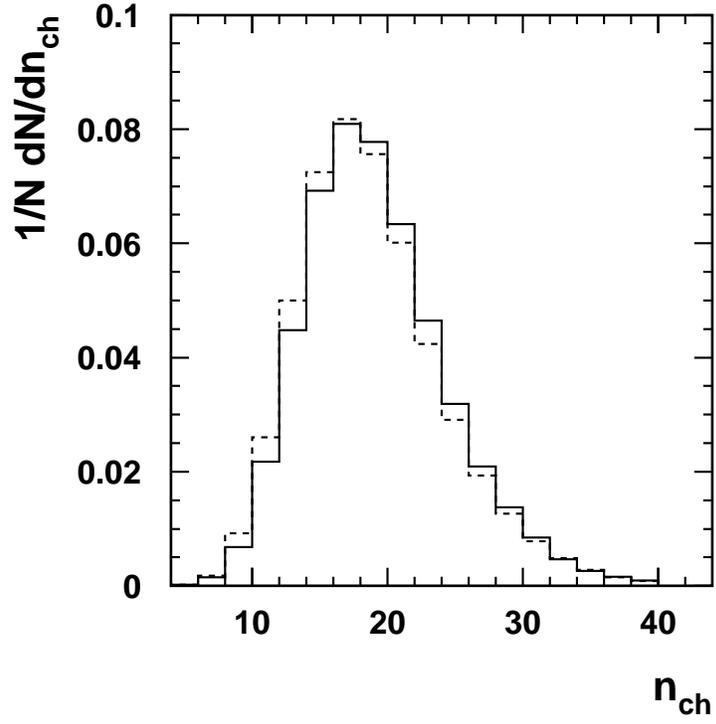}
                                                                                
\caption{Multiplicity distribution of charged particles
in three-jet light-quark Y events. The range of angle 
$\theta$ is $[0^{\circ},120^{\circ}]$.
The solid line is from CC events and the dashed line
is from CS ones. }
\label{fig.4}
\end{figure}









\begin{figure}
                                                                                
\includegraphics[scale=0.5]{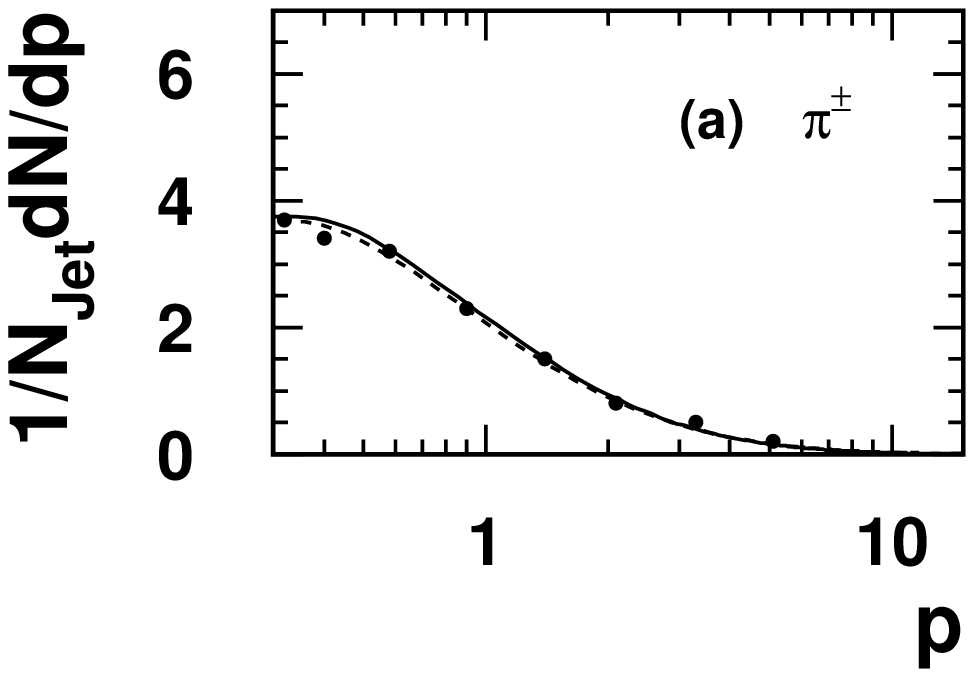}
\includegraphics[scale=0.52]{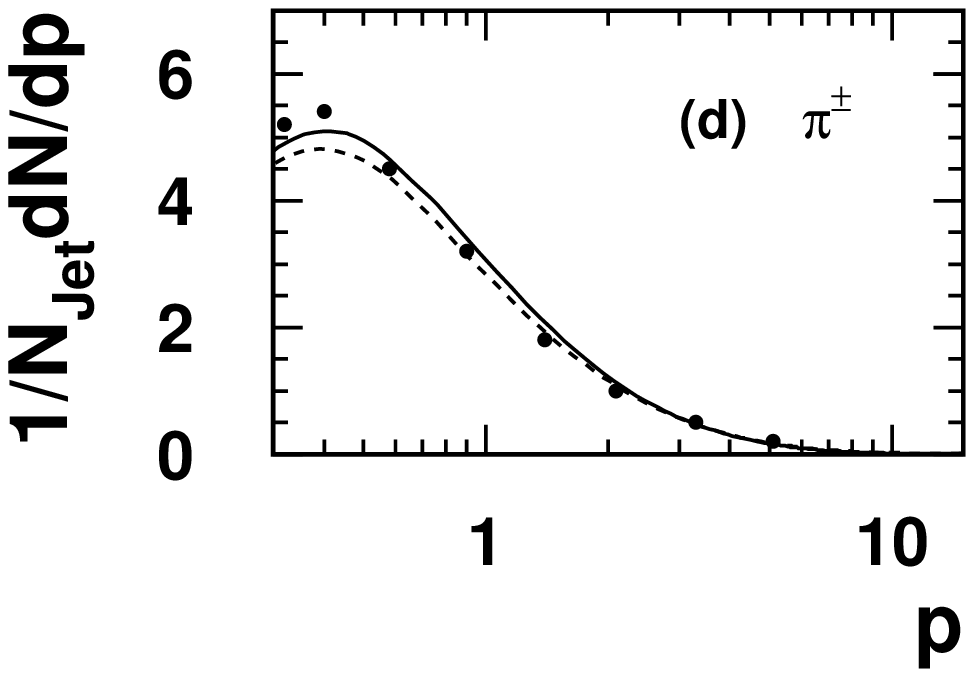}
\includegraphics[scale=0.5]{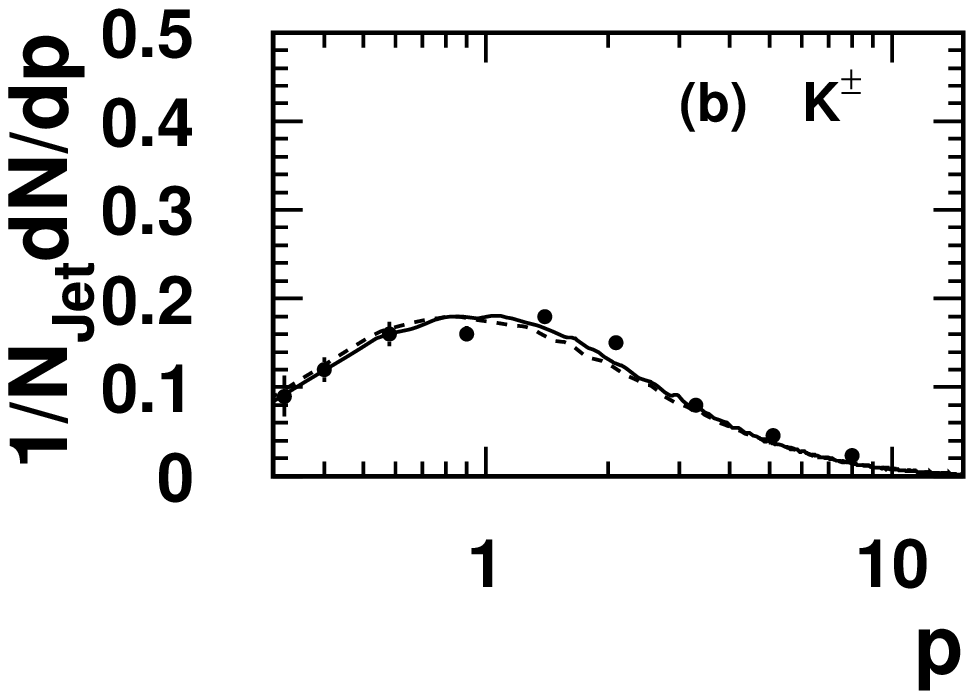}
\includegraphics[scale=0.5]{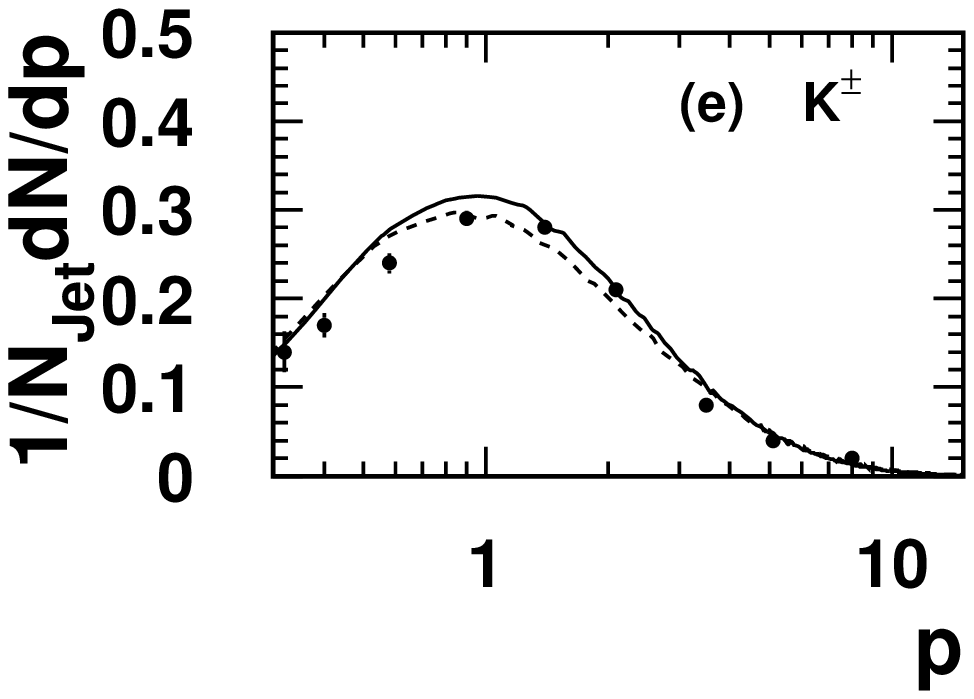}
\includegraphics[scale=0.5]{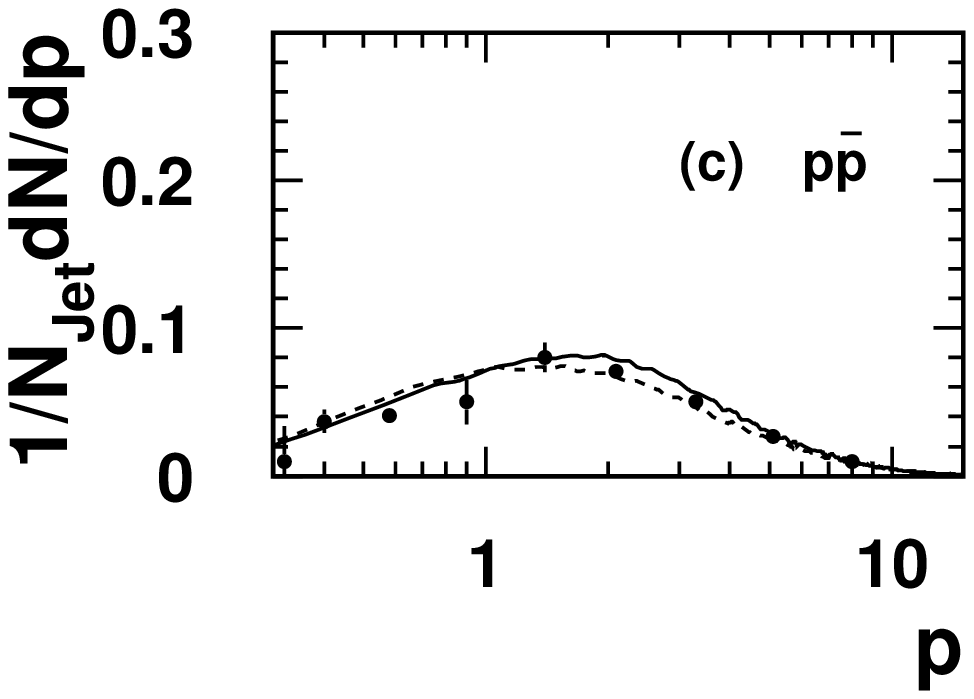}
\includegraphics[scale=0.5]{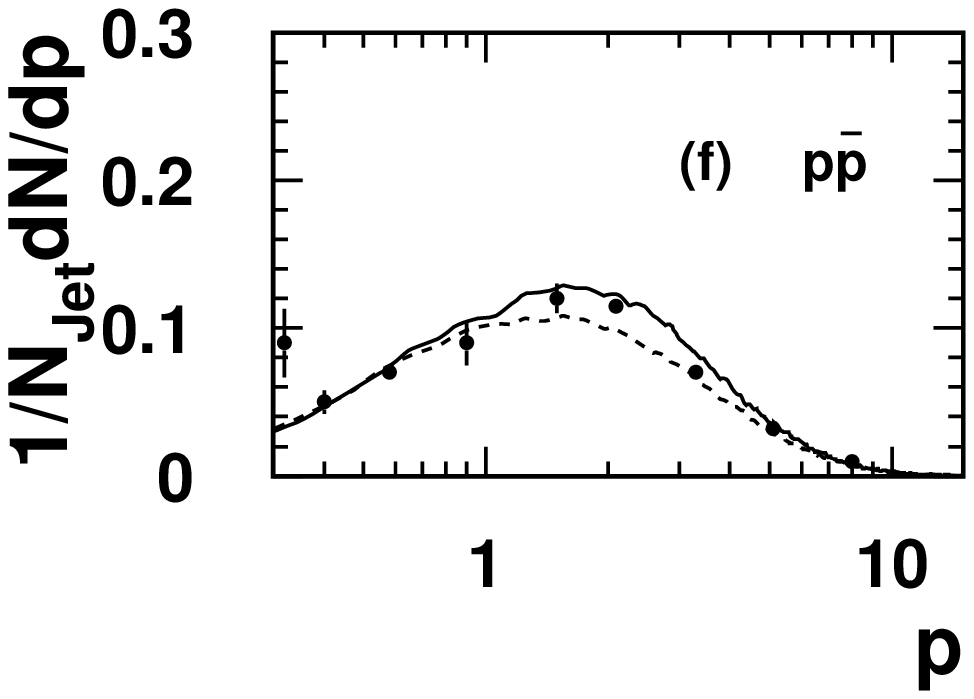}
                                                                                
\caption{Momentum spectra of identified hadrons in quark and gluon jets
for Y events. Figure (a) is spectra of pions, (b) kaons, and (c) protons 
in quark jets. (d), (e) and (f) are corresponding spectra in gluon jets. 
The solid line is from CC events and the dashed one is from CS events.
The data are taken from Ref. \cite{DELPHI2000}.
}
\label{figs.5(a)-(f)}
\end{figure}

\begin{figure}

\includegraphics[scale=0.54]{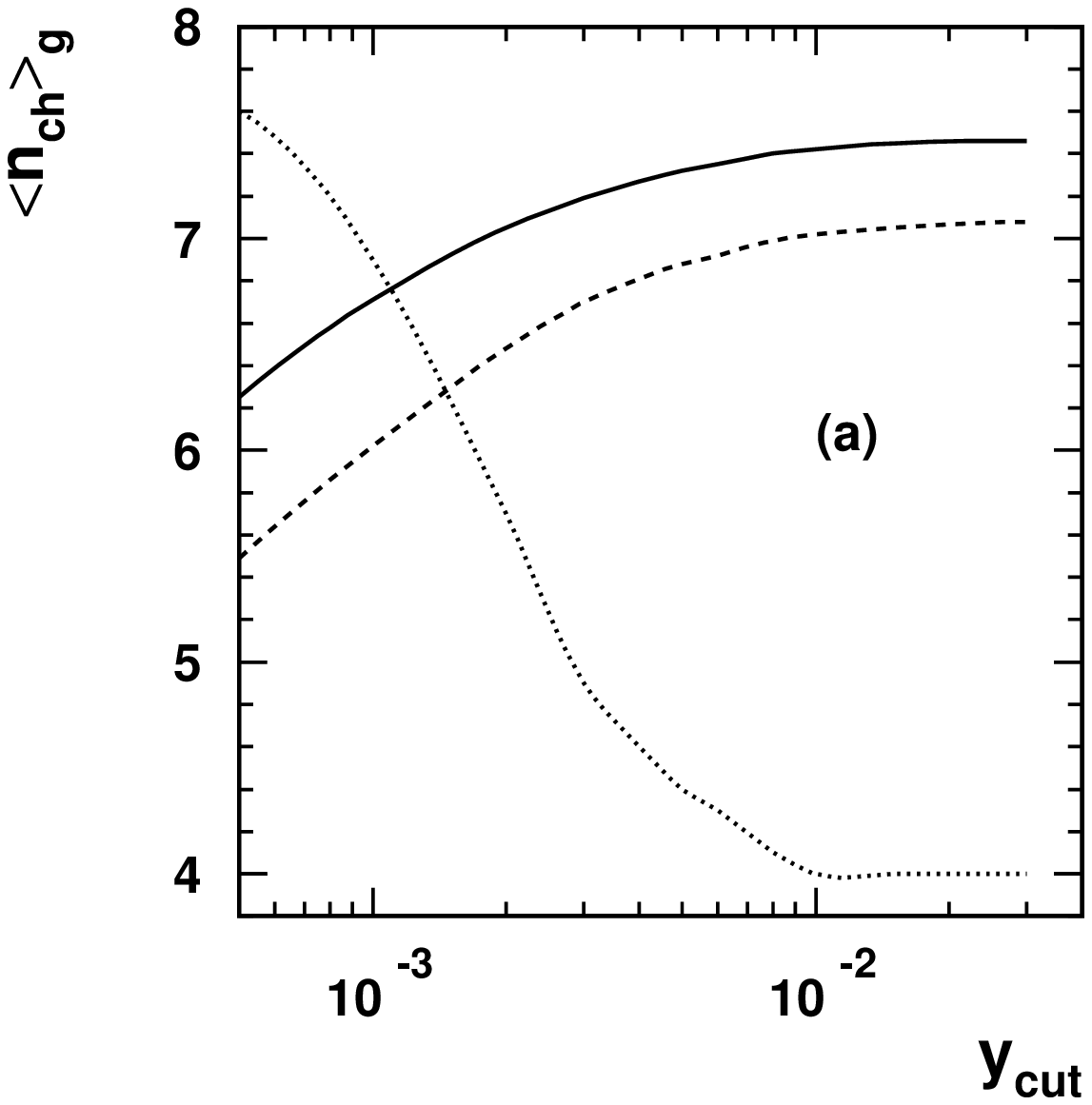}
\includegraphics[scale=0.54]{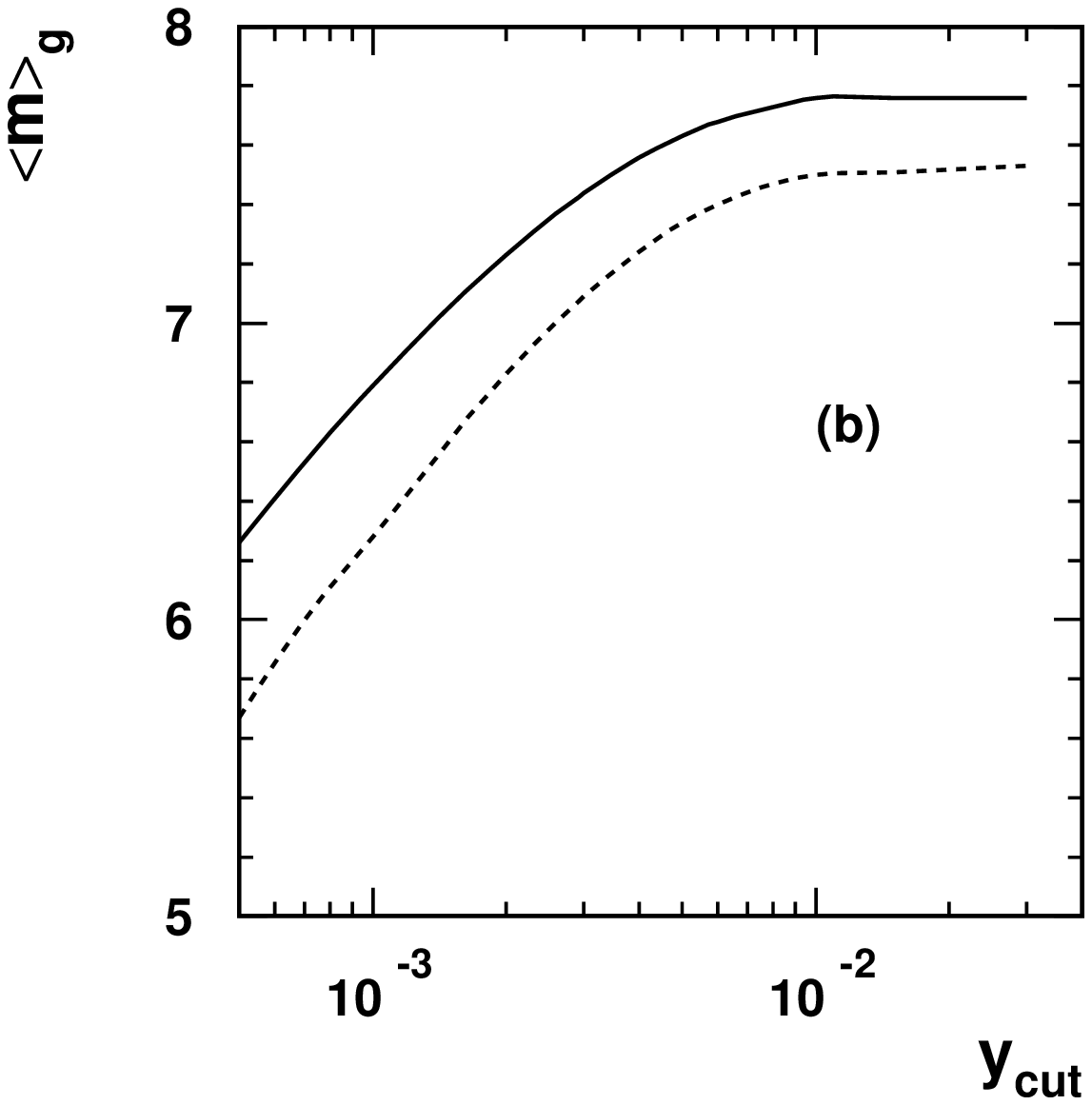}
\includegraphics[scale=0.54]{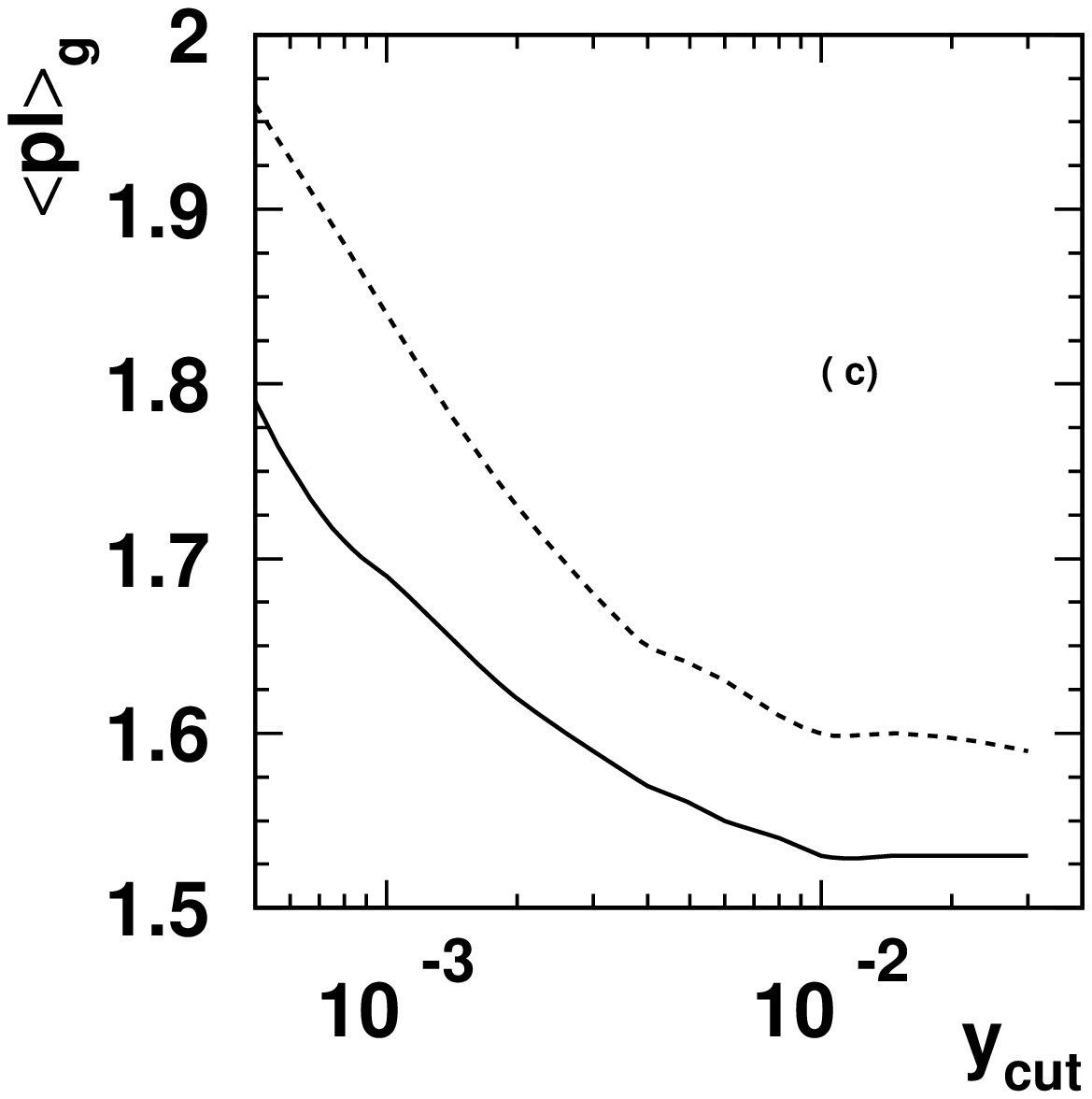}
                                                                                
\caption{The average values of observables for
the gluon-jet in three-jet events as functions
$y_{cut}$ at $Z^0$ pole.
$\theta_{12}$ and $\theta_{13}$ are limited within
$[100^{\circ},160^{\circ}]$.
(a) Multiplicity. The solid, dashed and dotted lines are
for CC, CS and their difference ($\times 10$) respectively;
(b) Invariant mass;
(c) Average longitudinal momentum. For both (b) and (c),
the solid and dashed lines are for CC and CS events respectively.
}
\label{figs.6(a)-(c)}
\end{figure}

\begin{figure}

\includegraphics[scale=0.8]{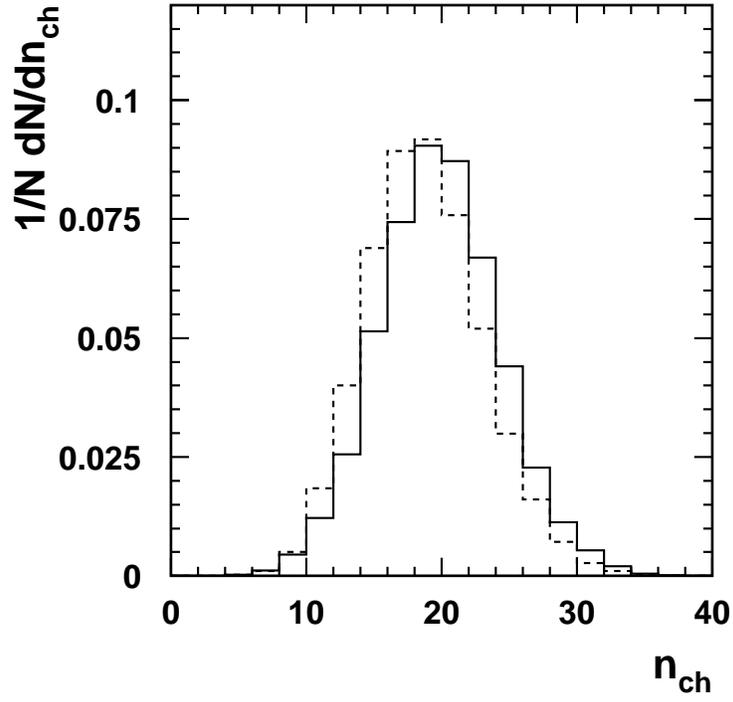}

\caption{The multiplicity distribution of charged particles 
in the selected three-jet events ($y_{cut}=0.0005$). $\theta_{12}$ and $\theta_{13}$ 
are limited within $[100^{\circ},160^{\circ}]$.
The solid and dotted lines are for the CC and CS events respectively. }
\label{fig.7}
\end{figure}

\begin{figure}

\includegraphics[scale=0.5]{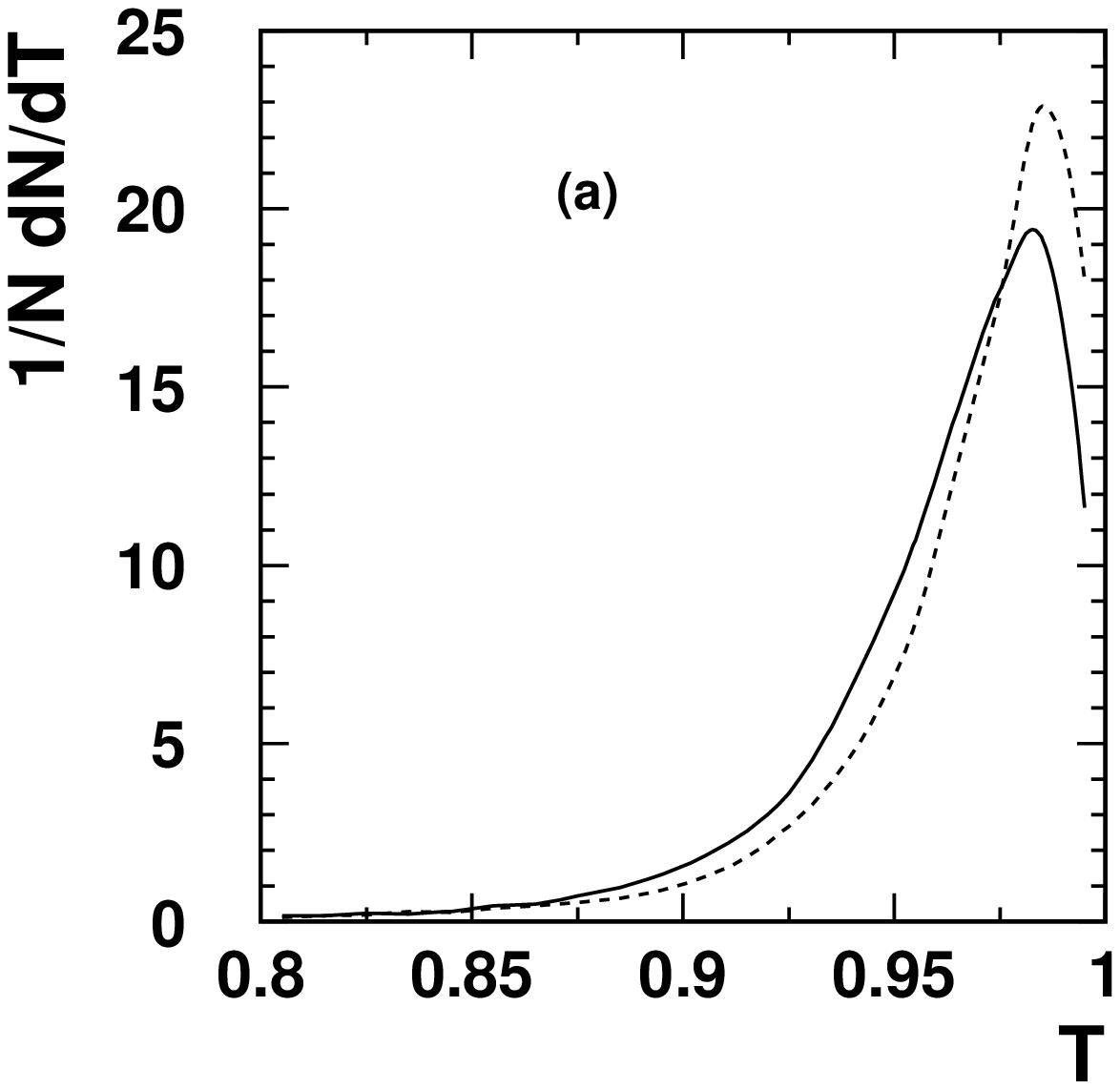}
\includegraphics[scale=0.52]{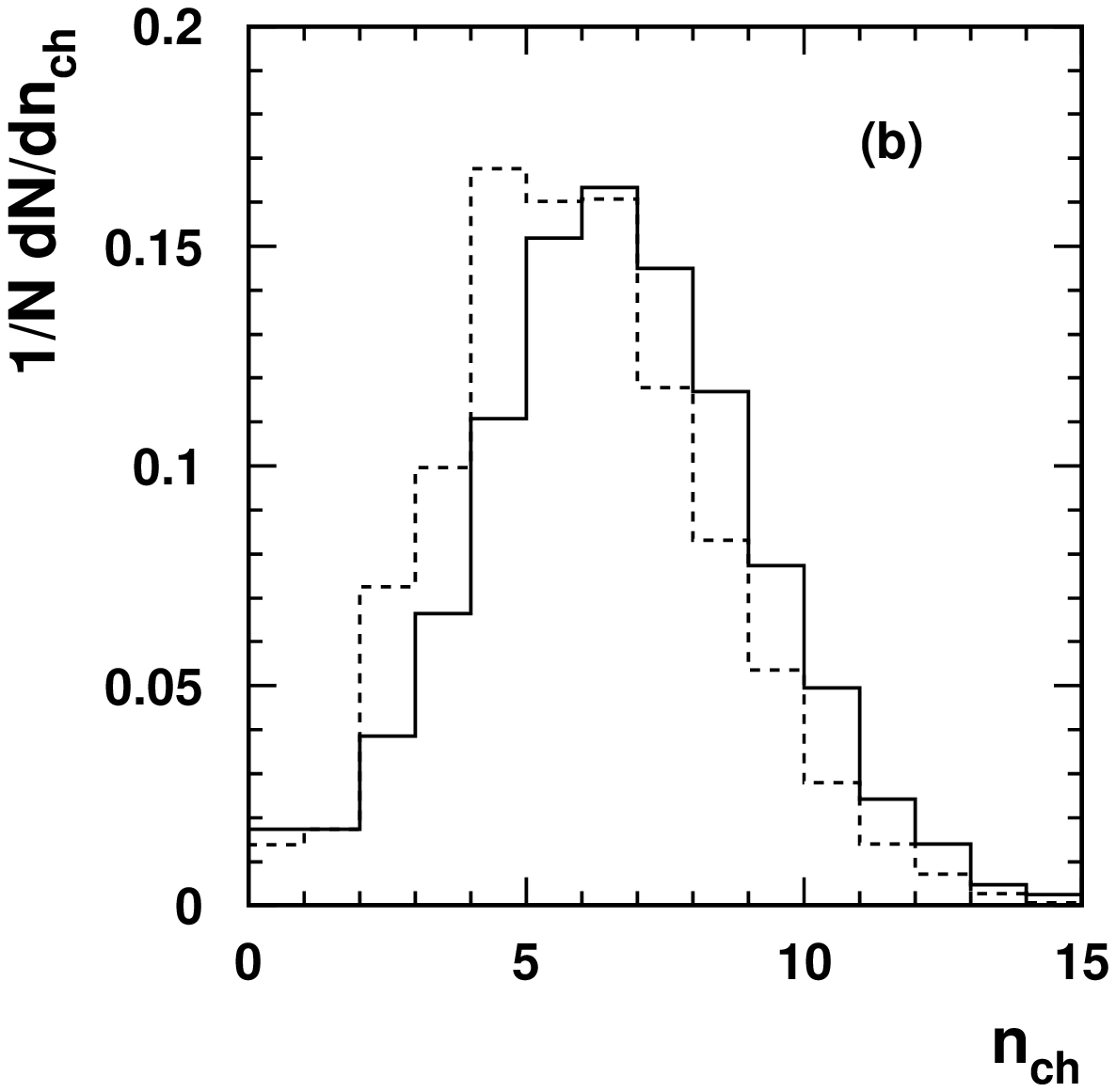}
\includegraphics[scale=0.5]{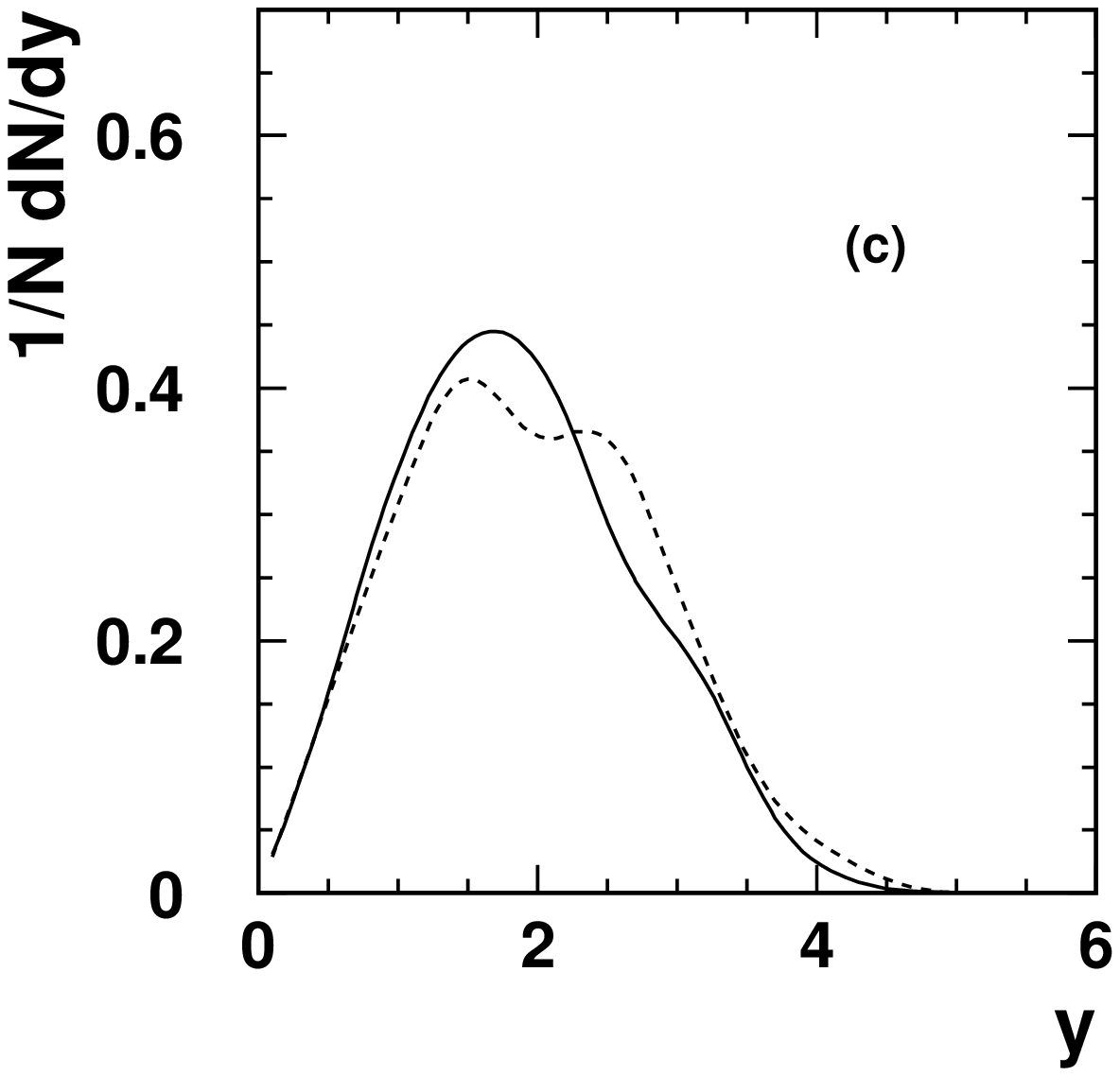}
\includegraphics[scale=0.5]{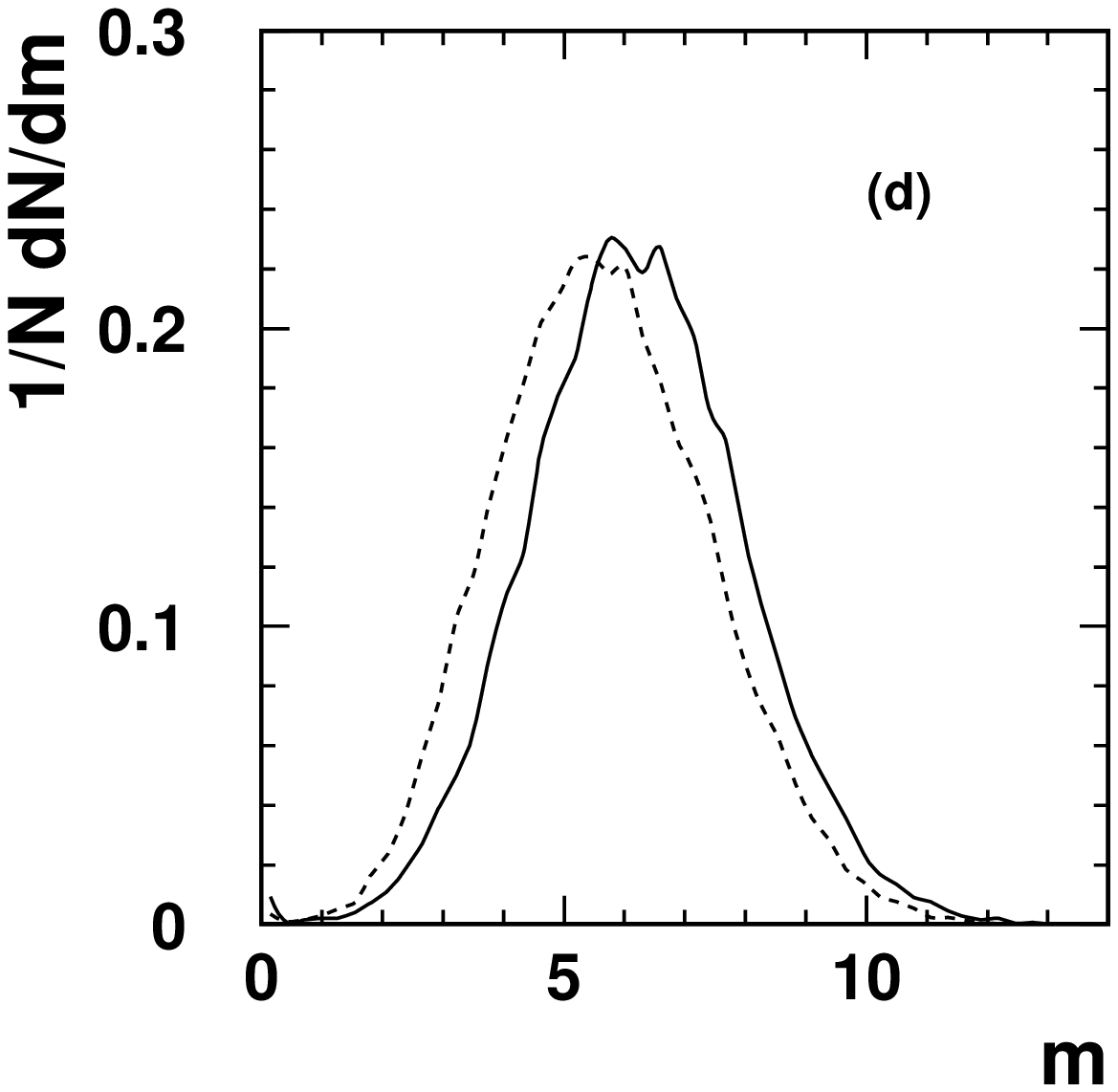}

\caption{Distributions of observables for the gluon jet in three-jet
events ($y_{cut}=0.0005$). $\theta_{12}$ and $\theta_{13}$ are limited within 
$[100^{\circ},160^{\circ}]$.
(a) T distribution;
(b) The distribution of the charged particle multiplicity;
(c) The rapidity distribution; (d) The invariant mass distribution.
The solid lines are for CC events, and the dashed lines are for
CS events.
}
\label{figs.8(a)-(d)}
\end{figure}

\end{document}